\documentclass[pra,superscriptaddress,twocolumn]{revtex4}

\usepackage{enumerate}
\usepackage{amsfonts,amssymb,amsmath}
\usepackage[]{graphics,graphicx,epsfig}
\usepackage{amsthm}

\usepackage{graphicx}
\usepackage{dcolumn}
\usepackage{natbib}
\usepackage{color}
\usepackage{multirow}

\def\identity{\leavevmode\hbox{\small1\kern-3.8pt\normalsize1}}

\newtheorem{propo}{Proposition}
\newcommand{\be}{\begin{eqnarray}}
\newcommand{\ee}{\end{eqnarray}}
\newcommand{\bpr}{\begin{propo}}
\newcommand{\epr}{\end{propo}}
\newcommand{\bpf}{\begin{proof}}
\newcommand{\epf}{\end{proof}}

\renewcommand{\epsilon}{\varepsilon}

\begin{document}

\title{Monogamy of particle statistics in tripartite systems simulating bosons and fermions}

\author{Marcin Karczewski}   
\affiliation{Faculty of Physics, Adam Mickiewicz University, Umultowska 85, 61-614 Pozna\'n, Poland}

\author{Dagomir Kaszlikowski}
\affiliation{Centre for Quantum Technologies,
National University of Singapore, 3 Science Drive 2, 117543 Singapore,
Singapore}
\affiliation{Department of Physics,
National University of Singapore, 3 Science Drive 2, 117543 Singapore,
Singapore}

\author{Pawe\l{} Kurzy\'nski}   \email{pawel.kurzynski@amu.edu.pl}   
\affiliation{Faculty of Physics, Adam Mickiewicz University, Umultowska 85, 61-614 Pozna\'n, Poland}
\affiliation{Centre for Quantum Technologies, National University of Singapore, 3 Science Drive 2, 117543 Singapore, Singapore}

\date{\today}


\begin{abstract}

In the quantum world correlations can take form of entanglement which is known to be monogamous. In this work we show that another type of correlations, indistinguishability, is also restricted by some form of monogamy. Namely, if particles A and B simulate bosons, then A and C cannot perfectly imitate fermions. Our main result consists in demonstrating to what extent it is possible.

\end{abstract}

\maketitle


\emph{Introduction.} If a particle in box A is indistinguishable from a particle in box B, then all the parameters describing their states, apart from their positions, are the same. Such particles are strongly correlated because the state of one of them automatically determines the state of the other. In quantum theory the above system is described by quantum correlations, i.e., by an entangled state which is either symmetric (bosons) or antisymmetric (fermions).       

Quantum entanglement cannot be efficiently explained within any reasonable classical theory \cite{Horodeccy}. Even within the quantum theory the description of entanglement is not simple if it occurs between more than two particles. Firstly, in these cases  entanglement can take many inequivalent forms \cite{Horodeccy,3ent,3ent2}. For instance, already for three particles there is more than one way to define a maximally entangled state. In addition, quantum correlations are restricted by the so called monogamy bounds \cite{mono1,CKW,mono2}. Specifically, if particles A and B are maximally entangled, then no other particle C can be entangled with either A or B. 

In multipartite systems indistinguishability can also take different forms. For example, three particles can be prepared in a tripartite symmetric state (three bosons), antisymmetric state (three fermions) or some general state allowing different pairs of particles to obey different statistics. It is intuitively clear that while particles in boxes A and B behave like bosons, it is not possible that particles in boxes B and C behave like fermions. This  could be interpreted as a manifestation of some form of monogamy.

However, in order to truly speak of a monogamy between different types of statistics one needs to consider a more detailed problem: given that particles in boxes A and B behave like bosons with probability $p$, what is the probability that particles in B and C behave like fermions? Here we answer the above question. More precisely, we derive tradeoff bounds on simulability of bosons and fermions in tripartite systems and represent them in a simple graphical way.

In this work we say that particles in a given state simulate some type of particle statistics, rather than that they are particles of a certain type (bosons, fermions, etc.). This is because in our considerations we assume that the same system can be prepared in states corresponding to different statistics. In general, this assumption requires application of asymmetric operations, which is only possible if the corresponding particles are distinguishable. Therefore, we prefer to say that here we discuss simulation of indistinguishability with entanglement.

In the following sections we introduce a class of tripartite systems whose properties are going to be studied in this work and recall basic features of permutation operators $\Pi_{XY}$. Their average values $v_{XY}=\langle \Pi_{XY} \rangle$ indicate if a pair of particles in modes $X$ and $Y$ exhibit perfect bosonic ($v_{XY}=1$), perfect fermionic ($v_{XY}=-1$), or any other type of behavior ($-1<v_{XY}<1$). Finally, we provide a tight monogamy relation for $v_{AB}$, $v_{BC}$ and $v_{AC}$ which can be used to further study general restrictions on simulability of bosonic and fermionic properties in multipartite systems.

Before we proceed, we would like to mention that our work is in a sense complementary to the recent one by Menssen {\it et. al.} \cite{menssen2017distinguishability}. There, the authors argued that knowing only the indistinguishability properties between pairs of particles is not enough to predict the effects resulting from indistinguishability of more than two particles. Here, we show that indistinguishability properties between the pair of particles impose some restrictions on indistinguishability properties of other pairs.




\emph{Preliminaries.} Let us consider a set of three modes $A$, $B$ and $C$ to which we refer as "boxes". Each box is occupied by a single particle, so that the general pure state of the system is of the form
\begin{eqnarray}\label{psi}
|\Psi\rangle &=& \alpha_1|A,B,C\rangle+\alpha_2|B,A,C\rangle+\alpha_3|C,A,B\rangle \nonumber \\
&+&\alpha_4|C,B,A\rangle+\alpha_5|A,C,B\rangle+\alpha_6|B,C,A\rangle.
\end{eqnarray}
Here we assume that the three particles are distinguishable, therefore the notation $|A,B,C\rangle$ means that the first particle is in box $A$, the second in $B$ and the third in $C$. Moreover, for simplicity we do not consider mixed states of the system. Nonetheless, our reasoning will apply to them as well. 

Next, we introduce  permutation operators $\Pi_{XY}$  swapping the labels of boxes $X$ and $Y$, for instance
\begin{eqnarray}
\Pi_{AB}|A,B,C\rangle =|B,A,C\rangle.
\end{eqnarray}
The average value of these operators, denoted as
\begin{eqnarray}
v_{XY} = \langle \Psi|\Pi_{XY}|\Psi\rangle, \label{v} 
\end{eqnarray}
 has a clear operational meaning. If $v_{XY}=\pm 1$ then the Hong-Ou-Mandel experiment \cite{Hong87} conducted on particles from boxes $X$ and $Y$ would result in perfect bunching/antibunching. Since bunching/antibunching is a typical bosonic/fermionic property that can be used as an indicator of particle statistics, in this work we say that if $v_{XY}=\pm1$, then the particles in boxes $X$ and $Y$ simulate bosons/fermions. In general, for $-1 < v_{XY} < 1$ the bunching/antibunching is not perfect and the particles in boxes $X$ and $Y$ simulate bosons/fermions with probability $\frac{1\pm v_{XY}}{2}$.

Let us also observe that a cyclic permutation operator $S$, defined as $S|X,\,Y,\,Z\rangle=|Y,\,Z,\,X\rangle$, can be expressed in terms of permutation operators $\Pi_{AB}$, $\Pi_{BC}$ and $\Pi_{AC}$
\begin{eqnarray}
S=\Pi_{AB}\Pi_{BC}&=&\Pi_{AC}\Pi_{AB}=\Pi_{BC}\Pi_{AC}. \label{abc}
\end{eqnarray}
Since $S^3=\openone$,  its eigenvalues are $1$ and $e^{\pm i\frac{2\pi}{3}}=-\frac{1}{2}\pm i\frac{\sqrt{3}}{2}$.


\emph{Monogamy between simulation of bosons and fermions.} Let us consider the relations between the permutation properties of different subsystems. We start with a simple example of $v_{AB}=1$, which means that the particles occupying boxes $A$ and $B$ simulate bosons. Then $\langle S \rangle=\langle \Psi|\Pi_{AB}\Pi_{BC}|\Psi\rangle=\langle \Psi|\Pi_{BC}|\Psi\rangle = v_{BC}$. Since the spectrum of $S$ is $1$ and  $-\frac{1}{2}\pm i\frac{\sqrt{3}}{2}$, the smallest real value attainable by $\langle S \rangle$ is $-\frac{1}{2}$. Therefore, maximal possible fermionic-like behavior in this case is given by $v_{BC}=-\frac{1}{2}$. As a result, $\langle S \rangle=-\frac{1}{2}$ and because of (\ref{abc}) $v_{AC}=-\frac{1}{2}$. An example of a state leading to the above values is
\begin{equation}
\frac{1}{2}\left(|A,B,C\rangle+|B,A,C\rangle-|A,C,B\rangle-|B,C,A\rangle\right).
\end{equation}

Similarly, one can ask about maximal bosonic behavior of $B$ and $C$, provided that $A$ and $B$ simulate fermions. In this case $v_{AB}=-1$ and $\langle S \rangle=-v_{BC}=-v_{AC}$. In order to maximize $v_{BC}$ we need to minimize $\langle S \rangle$, which is again $-\frac{1}{2}$. Therefore, the maximal bosonic behavior is $v_{BC}=v_{AC}=\frac{1}{2}$.  The possible corresponding state is of the form
\begin{equation}
\frac{1}{2}\left(|A,B,C\rangle-|B,A,C\rangle+|A,C,B\rangle-|B,C,A\rangle\right).
\end{equation}

Note, that if particles in $A$ and $B$ simulate bosons and so do particles in $B$ and $C$, then particles in $A$ and $C$ simulate bosons too. This is because in this case $\langle S \rangle=v_{AB}=v_{BC}=v_{AC}$. Therefore, simulability of bosons is transitive. This is also true for simulability of fermions in which case $-\langle S\rangle=v_{AB}=v_{BC}=v_{AC}$. 


\emph{General case.} The situation gets more complicated when neither pair of particles is perfectly bosonic or fermionic, i.e., $v_{AB}$, $v_{BC}$ and $v_{AC}$ are between $-1$ and $1$. To analyse this case we introduce the following three operators
\begin{eqnarray}
W_1 &=& \frac{1}{3}\left(\Pi_{AB}+\Pi_{BC}+\Pi_{AC}\right), \\
W_2 &=&  \frac{1}{3}\left(2\Pi_{AB}-\Pi_{BC}-\Pi_{AC}\right), \\
W_3 &=&  \frac{1}{\sqrt{3}}\left(\Pi_{BC} - \Pi_{AC}\right).
\end{eqnarray}
These operators have eigenvalues $\pm 1$ and $0$. For $W_1$ the eigenvalue $0$ is four times degenerate and the eigenvalue $+1$ corresponds to the symmetric state of the three particles (bosonic state), whereas $-1$ corresponds to the antisymmetric one (fermionic state). Interestingly, $W_1$ commutes with both $W_2$ and $W_3$, $[W_1,W_2]=[W_1,W_3]=0$, and in addition $W_1W_2=W_1W_3=0$.

There are two interesting properties of $W_2$ and $W_3$. Firstly, the two operators anticommute $\{W_2,W_3\}=W_2W_3+W_3W_2=0$. Secondly, $W_2^2 = W_3^2$. Because of these two properties the following operator
\begin{equation}
W_{\theta}=  W_2\,\cos\theta +  W_3\,\sin\theta,
\end{equation}
where $\theta\in [0,2\pi)$, obeys
\begin{eqnarray}
W_{\theta}^2 &=&  W_2^2\, \cos^2 \theta+  W_3^2\,\sin^2 \theta + \{W_2,W_3\} \,\cos \theta \,\sin\theta= \nonumber \\
&=& W_2^2 = W_3^2.
\end{eqnarray}
Note, that the above resembles the properties of Pauli spin-1/2 operators. The anticommutation of Pauli operators lies at the heart of the Bloch vector representation of spin-1/2 states \cite{NielsenChuang} and we will see in a moment that the properties of $W_2$   and $W_3$ also allow to propose a simple graphical representation of constraints on $v_{AB}$, $v_{BC}$ and $v_{AC}$.

The operator $W_1$ is supported on the subspace that is orthogonal to the one on which $W_2$ and $W_3$ are supported. Therefore
\begin{equation}\label{trade}
|\langle W_1 \rangle| + |\langle W_{\theta}\rangle| \leq 1,
\end{equation}
or explicitly
\begin{eqnarray}
& &|v_{AB}+v_{BC}+v_{AC}| + \label{tradeoff} \\
& &|(2v_{AB}-v_{BC}-v_{AC})\cos\theta +\sqrt{3}\, (v_{BC}-v_{AC})\sin\theta\, |\leq 3. \nonumber
\end{eqnarray}
The above constitutes a family of monogamy relations for $v_{AB}$, $v_{BC}$ and $v_{AC}$ parametrized by $\theta$. Any quantum state of the form (\ref{psi}) satisfies these relations for all $\theta$. 

The graphical representation of the relations (\ref{trade}) and (\ref{tradeoff}) defines a region in a three-dimensional space. With each state of form (\ref{psi}) (or a mixture of such states) one can associate the vector $\mathbf{v}=(v_{AB},v_{BC},v_{AC})$ which needs to lie inside this region. In order to find its shape let us first give a new representation of operators $W_1$, $W_2$ and $W_3$. We define a vector of operators
\begin{equation}
\mathbf{\Pi}=(\Pi_{AB},\Pi_{BC},\Pi_{AC}), 
\end{equation}
and  real vectors $\mathbf{w}_i$ ($i=1,2,3,\theta$) such that
\begin{eqnarray}
W_i=\mathbf{w}_i \cdot \mathbf{\Pi}.
\end{eqnarray}

\begin{figure}[t]
\includegraphics[width=0.45 \textwidth]{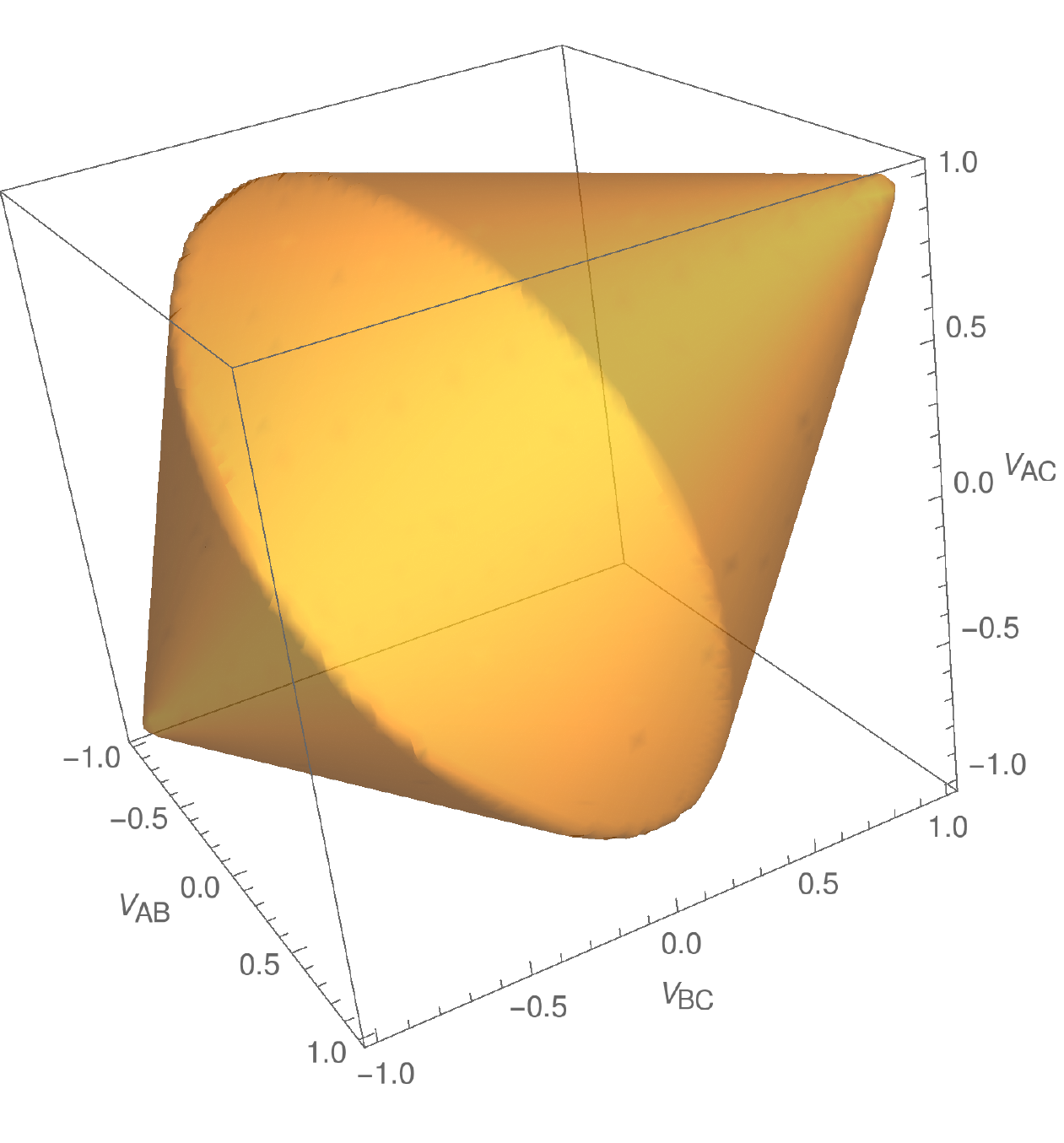}
\caption{Graphical representation of the treadoff relation (\ref{vectrade}). The vector $\mathbf{v}=(v_{AB},v_{BC},v_{AC})$ lies inside the region. }
\label{fig1}
\end{figure}

The average value of $W_i$ can then be represented as
\begin{equation}
\langle W_i \rangle = \mathbf{w}_i \cdot \mathbf{v}.
\end{equation}
Therefore, the relations (\ref{trade}) take form
\begin{equation}\label{vectrade}
| \mathbf{w}_1 \cdot \mathbf{v}|+| \mathbf{w}_{\theta} \cdot \mathbf{v}| \leq 1.
\end{equation}
The above formula defines two conical regions whose circular bases are connected (see Fig. \ref{fig1}). The bases lie in the plane spanned by $\mathbf{w}_2$ and $\mathbf{w}_3$ and the whole region has a rotational symmetry with respect to the $\mathbf{w}_1$ axis.  

Because of circular symmetry one can get rid of the parameter $\theta$ and write a single monogamy relation
\begin{equation}\label{tradesqrt}
| \mathbf{w}_1 \cdot \mathbf{v}|+ \sqrt{ (\mathbf{w}_{2} \cdot \mathbf{v})^2 + (\mathbf{w}_{3} \cdot \mathbf{v})^2 } \leq 1.
\end{equation}
However, the parameter $\theta$ allows one to show that the above monogamy relation is tight, i.e., one can find a quantum state corresponding to each point on the surface of the region denoted by (\ref{vectrade}).

Consider the state
\begin{equation}\label{surface}
|\chi^{(\pm,\pm)}_{\theta,\varphi}\rangle = \cos\varphi |\pm\rangle + \sin\varphi |\Psi_{\theta}^{\pm}\rangle, 
\end{equation}
such that 
\begin{eqnarray}
\langle {\pm}|W_1|{\pm}\rangle = \pm 1, \\
\langle \Psi_{\theta}^{\pm}|W_{\theta}|\Psi_{\theta}^{\pm}\rangle = \pm 1.
\end{eqnarray}
In the above $|\pm\rangle$ correspond to the symmetric (bosonic) and antisymmetric (fermionic) states of the three particles and $|\Psi_{\theta}^{\pm}\rangle$ are the $\pm 1$ eigenstates of the operator $W_{\theta}$. Because $W_1$ and $W_{\theta}$ span orthogonal subspaces we get $\langle {\pm}|\Psi_{\theta}^{\pm}\rangle =0$ and therefore
\begin{equation}
|\langle W_1 \rangle| + |\langle W_{\theta}\rangle|=|\cos\varphi|^2+|\sin\varphi|^2=1.
\end{equation}
The states (\ref{surface}) cover the whole surface for $\theta \in [0,\pi )$ and $\varphi \in [0,\pi/2]$.


\emph{Simulability of imperfect bosons and fermions is not transitive.} Now we discuss properties of a particular state. Let us consider a state corresponding to the vector $\mathbf{v}=(x,x,-x)$. We are looking for the maximal $x$ for which $\mathbf{v}$ satisfies the monogamy relation. The inequality (\ref{tradesqrt}) implies that the maximal value is $x=\frac{3}{5}$. The corresponding state is $\sqrt{\frac{4}{5}}|\phi\rangle + \frac{1}{\sqrt{5}}|+\rangle$ which is a superposition of the symmetric (bosonic) state and 
\begin{equation}
|\phi\rangle=\frac{1}{2}(|A,B,C\rangle+|B,A,C\rangle-|C,B,A\rangle-|B,C,A\rangle).
\end{equation}

This example shows substantial departure from transitivity of simulability of imperfect bosons and fermions. Although a pair of particles in boxes $A$ and $B$ (or $B$ and $C$) simulates bosons with the probability $\frac{1+v_{AB}}{2}=\frac{4}{5}$, the pair in boxes $A$ and $C$ simulates fermions with the probability $\frac{1-v_{AC}}{2}=\frac{4}{5}$. Note that the probability of bunching greater than $\frac{3}{4}$ is a sufficient condition to violate the noncontextuality-like inequality \cite{BosonicContext1,BosonicContext2,BosonicContext3}. The violation of this inequality implies that bunching properties of particles cannot be explained by a classical hidden variable model for which it is predetermined whether the particle goes through or reflects from a beam splitter.

Because the monogamy relation (\ref{vectrade}) is symmetric under the transformation $\mathbf{v}\rightarrow -\mathbf{v}$, a state for which $\mathbf{v}=\left(-\frac{3}{5},-\frac{3}{5},\frac{3}{5}\right)$ exists as well.


\emph{Glance at the four-partite scenario.} We now turn our attention to the case of four particles in four boxes, denoted by letters from A to D. By analogy with the three box scenario we consider six $v_{XY}$ parameters corresponding to average values of permutations $\Pi_{XY}$. In particular, we ask how does the minimal $v_{XY}$ change when some pairs are set to simulate bosons.

\begin{figure}[t]
\includegraphics[width=0.5 \textwidth]{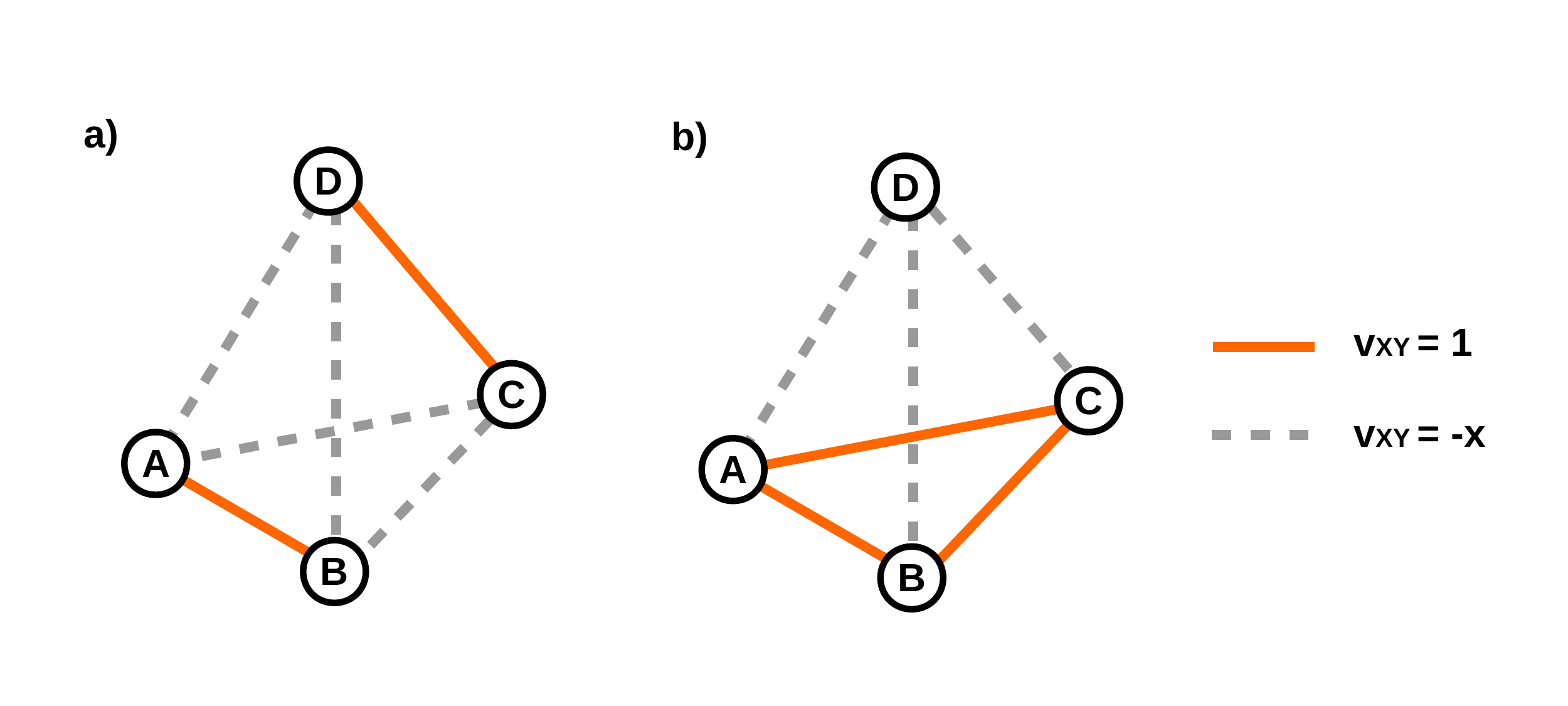}
\caption{Two examples of four-partite scenarios. The solid edges indicate simulation of bosons ($v_{XY}=1$) and the dashed edges denote simulation of imperfect fermions ($v_{XY}=-x,\,0<x<1$). The tripartite treadoff relations (\ref{vectrade}) can be applied to such systems by considering subsets of all particles. In case a) these relations are enough to derive the maximal value of $x$, but in case b) they need to be supplemented with additional fourpartite relations.}
\label{fig2}
\end{figure}

Figure \ref{fig2} presents two scenarios with significantly different properties. In the case a), $v_{AB}$ and $v_{CD}$ are set to 1 and we ask what is the minimal value of
\begin{equation}
-x=v_{AC}=v_{AD}=v_{BC}=v_{BD},~~0<x<1.
\end{equation}
The tripartite monogamy relations stemming from triangles ABC and ACD imply that $-x\geq - \frac{1}{2}$. One can show that this bound  is attained by the eigenstate corresponding to the largest eigenvalue of the operator
\begin{equation}
\Pi_{AB}+\Pi_{CD}-\Pi_{AC}-\Pi_{AD}-\Pi_{BC}-\Pi_{BD}.
\end{equation}
On the other hand, it is not possible to fully explain scenario b) with only tripartite relations. Once again, they imply that $-x\geq - \frac{1}{2}$. This time, however, $-x$ cannot be smaller than $-\frac{1}{3}$ as
\begin{equation}
\langle \Pi_{AB}+\Pi_{AC}+\Pi_{BC}-\Pi_{AD}-\Pi_{BD}-\Pi_{CD}\rangle=3+3x
\end{equation}
and the largest eigenvalue of the above operator is 4.


\emph{Summary and outlook.} Entangled particles can simulate indistinguishable particles, provided they are prepared in a proper state. If the state is symmetric,  the particles behave like bosons and if it is antisymmetric they behave like fermions. In general the particles can simulate various combinations of imperfect bosonic and fermionic properties. Here we show that these combinations are restricted by fundamental bounds. These bounds take form of monogamy relations, a tripartite version of which was derived in this work. Our relations are tight and can be represented using a simple three-dimensional visualization. One of the conclusions that stems from the relations is that simulability of imperfect bosons or fermions need not to be objective. Instead, it can be relative, i.e., the particle in box $A$ can simulate an imperfect boson when considered together with particle in box $B$ and at the same time an imperfect fermion when considered together with particle in box $C$. Finally, the tripartite relations can only partially explain the behavior of four-partite systems, which implies that indistinguishability, similarly to entanglement, depends on the number of particles.

We considered permutation properties of all three particle-pairs originating from a tripartite system. These properties can be related to the probabilities of bunching and anti-bunching in the Hong-Ou-Mandel experiment. However, following Menssen {\it et. al.} \citep{menssen2017distinguishability}, it would be interesting to consider genuine tripartite relations of the states studied in this work. For example, one can study the counting statistics for the states (\ref{psi}) on a symmetric three-port (tritter).

\emph{Acknowledgements.} M.K. and P.K. were supported by the National Science Centre in Poland through NCN Grant No. 2014/14/E/ST2/00585. D.K. was supported by the National Research Foundation and Ministry of Education in Singapore.




\end{document}